# Design and Implementation of DC-DC Buck Converter based on Deep Neural Network Sliding Mode Control

Zhi-Wei Liu, Wang-Bing Yu

School of Physics and Electronics, Hunan Normal University, Changsha, 410081, China

**Abstract:** In order to address the challenge of traditional sliding mode controllers struggling to balance between suppressing system jitter and accelerating convergence speed, a deep neural network (DNN)-based sliding mode control strategy is proposed in this paper. The strategy achieves dynamic adjustment of parameters by modelling and learning the system through deep neural networks, which suppresses the system jitter while ensuring the convergence speed of the system. To demonstrate the stability of the system, a Lyapunov function is designed to prove the stability of the mathematical model of the DNN-based sliding mode control strategy for DC-DC buck switching power supply. We adopt a double closed-loop control mode to combine the sliding mode control of the voltage inner loop with the PI control of the current outer loop. Simultaneously, The DNN performance is evaluated through simulation and hardware experiments and compared with conventional control methods. The results demonstrate that the sliding mode controller based on the DNN exhibits faster system convergence speed, enhanced jitter suppression capability, and greater robustness.

# 1. Introduction

Given that the majority of renewable energy sources generate DC power, it becomes imperative to convert DC voltage into various voltage levels using DC-DC converters. These converters are widely used in modern power technology. DC-DC buck switching power supplies serve as DC to DC step-down converters, commonly employed in scenarios requiring lower DC voltage supply. Typical applications include electric trolleys [1], microgrid systems [2], and DC motor equipment [3]. Nonetheless, the performance of buck switching power supplies is vulnerable to parameter deviations or external interference, thereby impacting the stability of their output voltage [4,5]. In practical applications, designing converters with excellent control strategies is crucial to enhancing the operational performance of buck-type switching power supplies.

In recent years, various advanced control techniques for DC-DC converters have been widely developed and applied. Kobaku [6] systematically designed a robust PID controller that uses only the detected output voltage as feedback. They applied quantitative feedback theory to address issues of uncertainty and unstable external perturbations in DC-DC converters. Li [7] proposed a multi objective adaptive switching control that enables fast switching of the DC-DC converter and enhances its smooth transient response during the recovery process. For resistive load perturbations and input voltage variations in DC-DC boost converters, Rong [8] designed a twisting sliding mode control for DC-DC boost converter using nonlinear disturbance observer. This method allows the system's output voltage to quickly and accurately track a given voltage, thereby enhancing tracking performance and robustness. Additionally, other control strategies, such as sliding mode control (SMC) [9,10], fuzzy control [11], and stepping control [12], have also been applied to enhance the stability of switching power supplies. SMC, a widely utilized control technique that has evolved significantly in recent years, has progressed from early first-order SMC to more recent second-order SMC [13]. Second-order SMC enhances performance metrics such as transient response, albeit requiring an additional capacitive current sensor for implementation. In the past, researchers using hysteresis-modulated (HM) SMC to achieve control of the DC-DC converter, but this approach resulted in a non-constant switching frequency. To address this issue, researchers have now developed PWM-based SMC, which ensures a consistent switching frequency. Wang [14] designed continuous nonsingular terminal sliding mode control by combining a disturbance observer with a PWM-based sliding mode method, improving voltage tracking performance. However, PWM-based SMCs usually have high switching frequency and sampling rate requirements, which can cause excessive losses and are not suitable for high-power devices. Most research on control strategies for switching power supplies primarily focuses on conventional control strategies, where parameters are derived from calculations and cannot be dynamically adjusted.

With the advancement of artificial intelligence technology, neural networks (NNs) are increasingly employed in the control of DC-DC converters to facilitate the adjustment of dynamic parameters. Lai [15] designed a neural network-based DC-DC analog-digital converter for underwater solar charging. Liu [16] designed a data-mining-based hardware-efficient neural network controller for DC–DC switching converters. Rojas-Dueñas [17] designed a wavelet convolutional neural network-based black-box model of a DC-DC converter to improve system performance. Deep neural networks (DNNs) are an important branch of neural networks, and although extensive research has been carried out on DNN-based optimal control of nonlinear systems [18-20], their applications in DC-DC converters are still rarely reported.

In this paper, the research focuses on the buck-type DC-DC switching power supply, with the introduction of DNN alongside the sliding mode control strategy. This integration enables the learning of the control strategy's output and facilitates online parameter adjustment. The entire process is controlled with digital precision, and the stability of the control strategy is validated using the Lyapunov function. Based on the simulation and experimental results, the DNN-based sliding mode control buck-type DC-DC converter designed in this paper consistently meets expectations in terms of output voltage, interference immunity, chattering, and convergence speed, even under varying load or input voltage conditions.

## 2. Methodology

*2.1 DC-DC buck converter*

The basic circuit for buck converter is shown in Fig. 1. The buck converter consists of a controlled switch $M$, diode $D$, capacitors $C$, inductors $L$, and a controlled drive load circuit $R$. $V_{in}$ denotes the input voltage and $V_{out}$ denotes the output voltage of the controlled drive load circuit. The switch controls the inflow and outflow of current by periodically turning on and off. The time that the switch is on during the entire cycle is called the duty cycle, and the duty cycle $D$ has a value from 0 to 1. When $D = 0$, zero voltage appears across the load; when $D = 1$, all input voltages appear across the load.

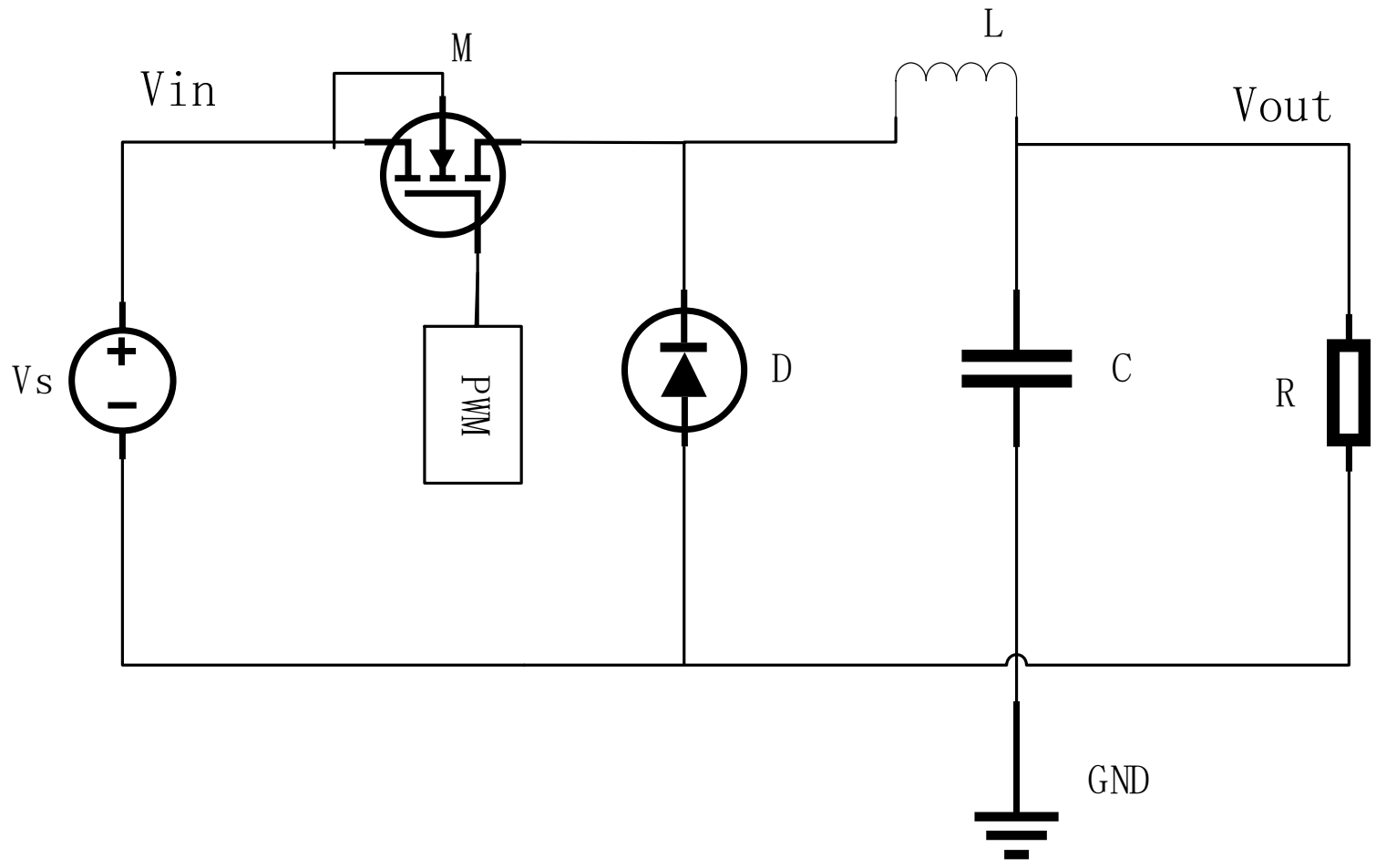

Fig. 1 Basic circuit for buck converter

Based on Kirchhoff's law, using $S_C = 0$ and $S_C = 1$ to denote the inflow and outflow, the differential equations of the circuit can be listed [21-23]: (1) for $S_C = 1$, transistor T is closed and diode D is open; (2) for $S_C = 0$, transistor T is closed and diode D is closed.

According to Kirchhoff's current and voltage laws, the buck DC-DC converter model is obtained as:

$$\begin{cases} \frac{di_l}{dt} = S_C \frac{V_{dc}}{L} - \frac{V_{load}}{L} \\ \frac{dV_{load}}{dt} = \frac{i_l}{C} - \frac{i_{load}}{C} \end{cases} \tag{1}$$

From the point of view of impedance loading, Eq. (1) becomes:

$$\begin{cases} \frac{di_l}{dt} = S_C \frac{V_{dc}}{L} - \frac{V_{load}}{L} \\ \frac{dV_{load}}{dt} = \frac{i_l}{C} - \frac{V_{load}}{RC} \end{cases} \tag{2}$$

According to state-space averaging, Eq. (2) can be written as the average dynamic model shown in Eq. (3):

$$\begin{cases} \frac{d\langle i_l \rangle_{T_S}}{dt} = \langle S_C \rangle_{T_S} \frac{\langle V_{dc} \rangle_{T_S}}{L} - \frac{\langle V_{load} \rangle_{T_S}}{L} \\ \frac{d\langle V_{load} \rangle_{T_S}}{dt} = \frac{\langle i_l \rangle_{T_S}}{C} - \frac{\langle V_{load} \rangle_{T_S}}{RC} \end{cases} \tag{3}$$

where $\langle i_l \rangle_{T_S}$, $\langle V_{load} \rangle_{T_S}$, $\langle S_C \rangle_{T_S}$ and $\langle V_{dc} \rangle_{T_S}$ are the average values of inductor current, output voltage, control signal, and input voltage, respectively, and $T_S$ is a conversion period. Eq. (3) can be simplified to a more concise form of an uncertain nonlinear system as shown in Eq. (4) below:

$$\dot{X} = f(X) + g(X)D \tag{4}$$

where $D$ denotes the duty cycle. The nonlinear equations $f(X)$ and $g(X)$, and the mean state vector $\dot{X}$ are defined as follows:

$$g(X) = \begin{pmatrix} \frac{V_{dc}}{L} \\ 0 \end{pmatrix} \tag{5}$$

$$f(X) = \begin{pmatrix} 0 & -\frac{V_{load}}{L} \\ \frac{i_l}{C} & -\frac{V_{load}}{RC} \end{pmatrix} \tag{6}$$

$$\dot{X} = \left[ \langle i_l \rangle_{T_S} \langle V_{load} \rangle_{T_S} \right]^T \tag{7}$$

Inductor current and capacitor voltage are selected as state variables $x_1$ and $x_2$. The mathematical model of the DC-DC buck switching power supply can be finally derived as:

$$\begin{bmatrix} x_1 \\ x_2 \end{bmatrix} = \begin{bmatrix} 0 & -\frac{1}{L} \\ \frac{1}{C} & -\frac{1}{RC} \end{bmatrix} \begin{bmatrix} x_1 \\ x_2 \end{bmatrix} + \begin{bmatrix} \frac{V_i}{L} \\ 0 \end{bmatrix} D \tag{8}$$

$$y = \begin{bmatrix} 0 & 1 \end{bmatrix} \begin{bmatrix} x_1 \\ x_2 \end{bmatrix} \tag{9}$$

where $D$ is the control law, $V$ is the input voltage, and $R$, $L$, and $C$ are the resistance, inductance, and capacitance parameters of the circuit, respectively.

*2.2 Sliding Mode Control*

Given the nonlinearity and significant uncertainties inherent in the DC-DC buck converter, employing a sliding mode control method proves to be an effective strategy. This control method exploits the uncertainty and unmeasurable external disturbances with excellent robustness and theoretically provides good control over buck circuits. In this paper, the sliding mode control method is used to design buck-type DC-DC circuits, and Fig. 2 shows the block diagram of the sliding mode controller.

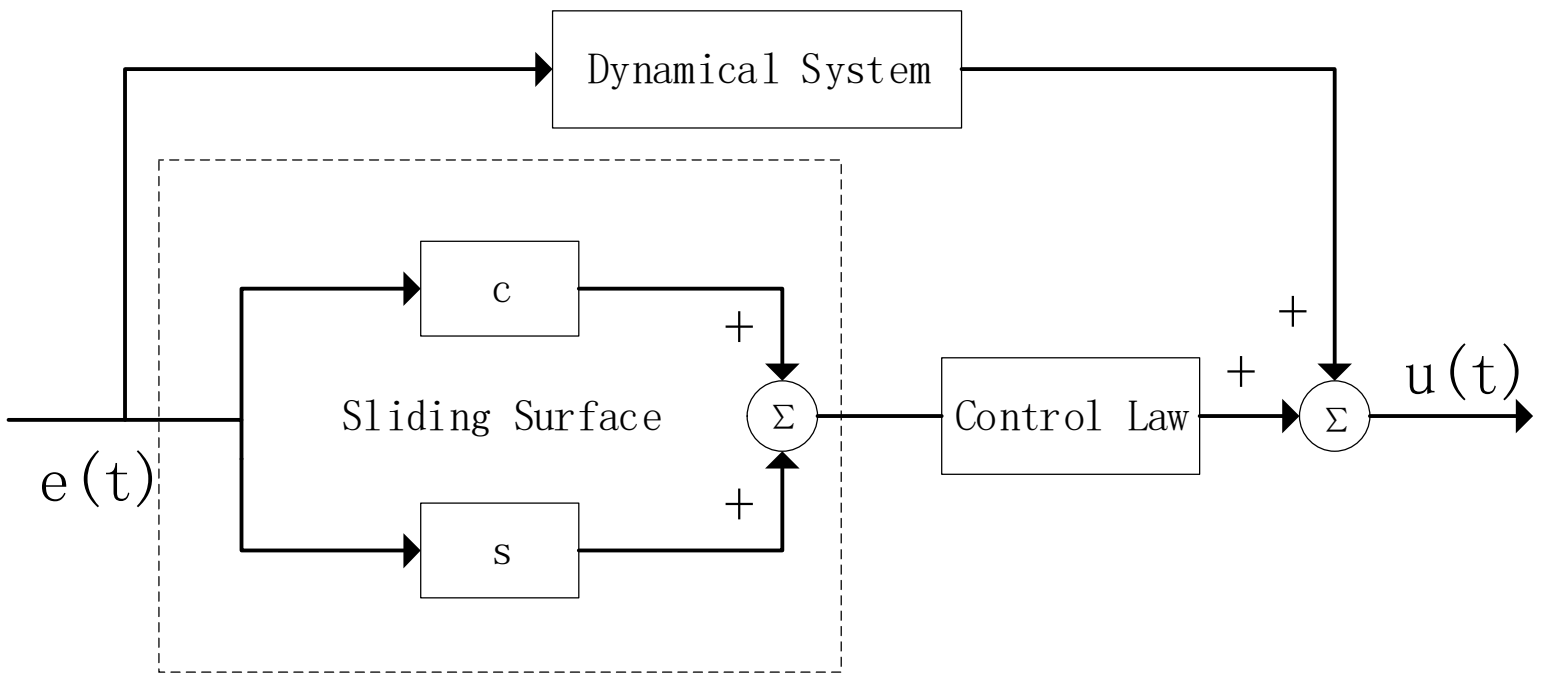

Fig. 2 The block diagram of the sliding mode controller

Principle of the sliding mode controller: introduce a hyperplane called the "sliding mode surface" and design the control law to make the system state slide rapidly on this plane, thereby achieving stability and robustness in the system. The choice of the sliding mode surface is crucial, as it should effectively resist external disturbances and system uncertainties. The goal of the control law is to guide the system state to slide rapidly on the sliding mode surface to suppress the effect of disturbances on the system. The characteristics of the entire system are determined by the designed sliding mode surface, making the system robust against external influences.

The proposed control system for sliding mode variable structure control is as follows:

$$\dot{x} = f(x, u, s) \quad x \in R^n, u \in R^m, t \in R \tag{10}$$

where $u$ is the control law, $s$ is time, and $x$ is the state variable.

Design a switching function for the system as follows:

$$s(x) \quad s \in R^m \tag{11}$$

Solve the system control rate function $u$ as follows:

$$u = \begin{cases} u^+(x) & s(x) > 0 \\ u^-(x) & s(x) < 0 \end{cases} \tag{12}$$

From (12), it is known that around the switching surface, the discontinuity in $u$ is the manifestation of the variable structure. The sliding mode variable structure needs to satisfy three conditions: (1) the existence of a sliding mode state; (2) stability of the Lyapunov function; and (3) in a finite period of time, the moving point can reach the vicinity of the sliding mode surface. Only if these three basic conditions are satisfied can it be called a successful design of the sliding mode variable structure control.

Based on the mathematical model of the DC-DC buck switching power supply introduced, the state space representation is transformed to incorporate the reference voltage $V_{ref}$ and the output voltage $V_o$. The difference between the reference voltage and the output voltage is selected as the new state variable $x_1$, while the derivative of $x_1$ is chosen as another state variable $x_2$. Consequently, the updated state space model can be expressed as follows:

$$x_1 = V_{ref} - V_o \tag{13}$$

$$x_2 = \dot{x}_1 = -\dot{V}_o \tag{14}$$

For DC-DC buck circuits, according to Ohm's law, KCL, and KVL, there are:

$$i_R = \frac{V_o}{R} \tag{15}$$

$$i_L = \int [(DV_i - V_o)/L]\, dt \tag{16}$$

And so it can be obtained:

$$\dot{x}_2 = -\ddot{V}_o = -\frac{1}{C}\frac{d(i_C - i_R)}{dt} = -\frac{1}{LC}x_1 - \frac{1}{RC}x_2 + \frac{V_{ref}}{LC} - \frac{V_i}{LC}D \tag{17}$$

The spatial model equation is obtained:

$$\begin{bmatrix} x_1 \\ x_2 \end{bmatrix} = \begin{bmatrix} 0 & 1 \\ -\frac{1}{LC} & -\frac{1}{RC} \end{bmatrix} \begin{bmatrix} x_1 \\ x_2 \end{bmatrix} + \begin{bmatrix} 0 \\ \frac{V_{ref}}{LC} \end{bmatrix} + \begin{bmatrix} 0 \\ -\frac{V_i}{LC} \end{bmatrix} D \tag{18}$$

Design sliding mode surface as:

$$s = cx_1 + \dot{x}_1 \quad c > 0 \tag{19}$$

where $c$ is the sliding mode surface parameter. Let $\dot{s}$, then:

$$\dot{s} = c\dot{x}_1 + \dot{x}_2 = cx_2 - \frac{1}{LC}x_1 - \frac{1}{RC}x_2 + \frac{V_{ref}}{LC} - \frac{V_i}{LC}D = 0 \tag{20}$$

Then the equivalent sliding mode control law can be expressed as:

$$u_{eq} = D = \frac{LC}{V_i}[cx_2 - \frac{1}{LC}x_1 - \frac{1}{RC}x_2 + \frac{V_{ref}}{LC}] \tag{21}$$

Due to the presence of various disturbances in the realistic system, the switching control term needs to be added. The expression for the switching control term of the designed system is given by using the isochronous convergence law:

$$u_{sw} = -\frac{LC}{V_i}\eta\, sgn(s) \quad \eta > 0 \tag{22}$$

The sliding mode control law consists of an equivalent sliding mode control as well as a system switching control:

$$u = u_{eq} + u_{sw} \tag{23}$$

$$u = \frac{LC}{V_i}[cx_2 - \frac{1}{LC}x_1 - \frac{1}{RC}x_2 + \frac{V_{ref}}{LC} - \eta\, sgn(s)] \tag{24}$$

Finally, the stability of the system is proved by the Lyapunov function. The Lyapunov function for the sliding mode control is:

$$V = \frac{1}{2}s^2 \tag{25}$$

Derivation of it:

$$\begin{aligned} \dot{V} &= s\dot{s} \\ &= s\left(cx_2 - \frac{1}{LC}x_1 - \frac{1}{RC}x_2 + \frac{V_{ref}}{LC} - \frac{V_i}{LC}D\right) \\ &= s\left(-\frac{LC}{V_i}\frac{V_i}{LC}\eta\, sgn(s)\right) \\ &= -\eta s\, sgn(s) \\ &= -\eta|s| \leq 0 \end{aligned} \tag{26}$$

Therefore, it is guaranteed that the sliding mode reaching condition holds.

# 3 DNN-based Sliding Mode Control

The DNN model consists of four layers. The first layer is the input layer, composed of four input nodes denoted as "$I_n$". The second and third layers are hidden layers, with the number of nodes determined by specific practical needs. Here, they are set to three nodes each, denoted as "$H_{an}$" and "$H_{bn}$". The fourth layer is the output layer, containing one output node denoted as "$O_n$". The weight parameters between "$I_n$" and "$H_{an}$", "$H_{an}$" and "$H_{bn}$", and "$H_{bn}$" and "$O_n$" are denoted by $m_1$, $m_2$, and $m_3$, respectively. The connection weights are denoted as $W = [m_1 + m_2 + m_3]^T$, as shown in Fig. 3.

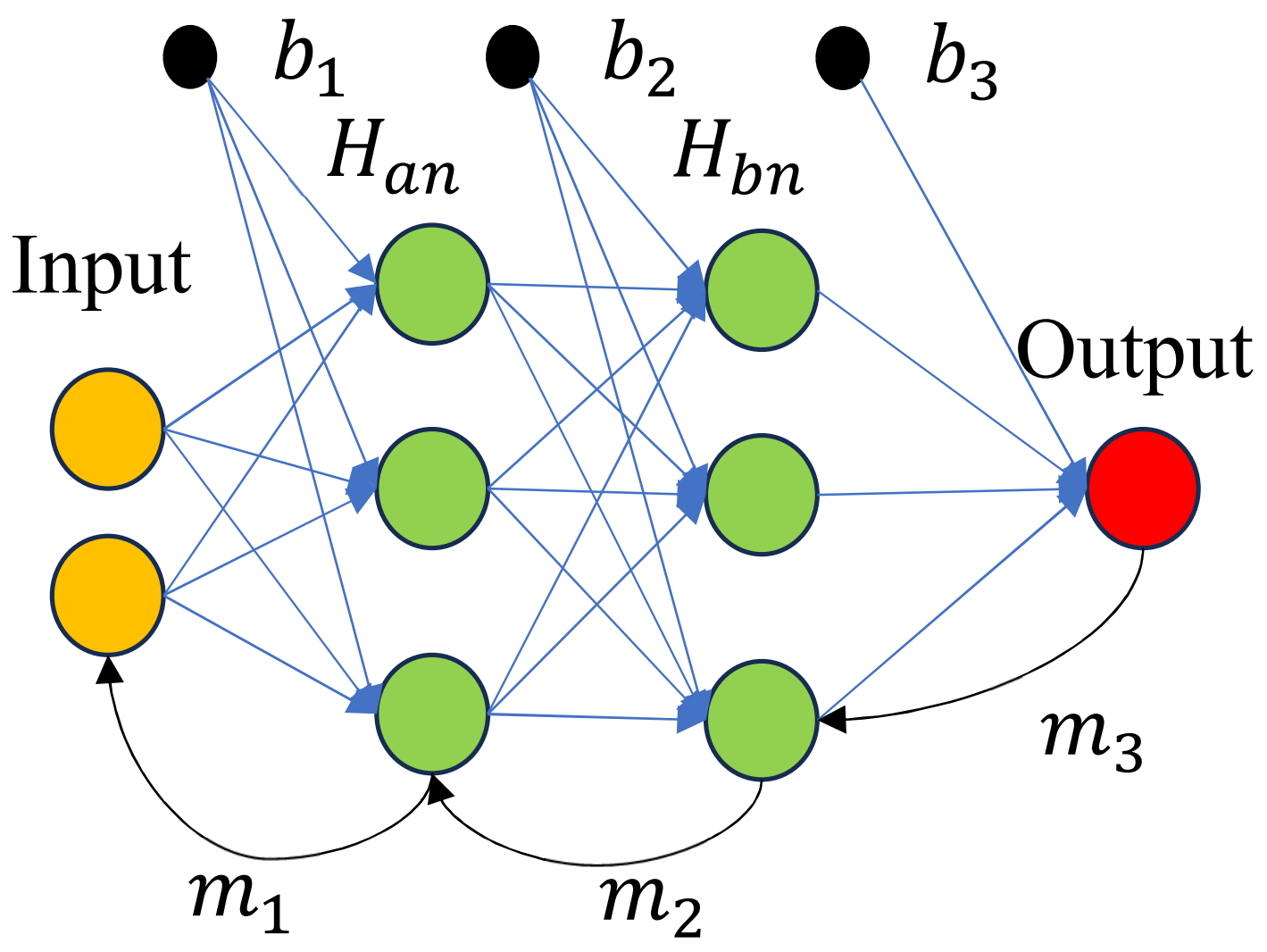

Fig. 3 DNN model

The Root Mean Square Error (RMSE) is a commonly used metric to evaluate the performance of a model. It measures the difference between predicted and actual values, whereupon the optimal combination of hyperparameters can be selected based on the RMSE. Selecting the best combination of optimization functions and activation functions is crucial for obtaining the best performance. Table 1 provides a list of RMSE values for different combinations of these techniques and functions. Each epoch contains a weight update, and the table shows a comprehensive overview of the performance of each combination at 50 epochs. By selecting the combinations with the smallest RMSE values, the optimal deep learning model and hyperparameters can be determined.

Table 1 Hyperparameters selection

| **Optimizer** | **Actitation Function** | | |
|---|---|---|---|
| | **Sigmoid** | **ReLU** | **Tanh** |
| **Adam** | 0.02256 | 0.00082 | 0.003549 |
| **RMSprop** | 0.03945 | 0.00652 | 0.00701 |
| **SGD** | 0.15299 | 0.00091 | 0.00752 |

Based on the data in Table 1, it can be concluded that the combination of the ReLU activation function with the SGD optimization function yields a lower RMSE compared to the other combinations. Furthermore, the ReLU function facilitates more efficient gradient descent and backpropagation methods compared to the other two functions, thereby helping to avoid issues such as gradient explosion and gradient vanishing.

The function realization of the DNN output is represented as:

$$\begin{cases} f(\boldsymbol{x}) = \boldsymbol{W}^{*T}\sigma(x) + \varepsilon \\ \sigma(\boldsymbol{x}) = \max(0, \boldsymbol{W}^T\boldsymbol{x} + \boldsymbol{b}) \end{cases} \tag{27}$$

where $\boldsymbol{W}^*$ represents the ideal weights; $f(\boldsymbol{x})$ is the output value and the real quantity that the DNN has to fit; $\varepsilon$ is the approximation error, which is a positive number $\varepsilon_N$ greater than zero.

The output is predicted to be:

$$S_1 = m_1 * I_n + b_1; H_{an} = f(S_1) \tag{28}$$

$$S_2 = m_1 * H_{an} + b_2; H_{bn} = f(S_2) \tag{29}$$

$$S_3 = m_3 * H_{bn} + b_3; O_n = f(S_3) = \widehat{O_\iota} \tag{30}$$

The cost function is utilized to minimize the RMSE between the predicted and output values, It is calculated as follows:

$$cost = \frac{1}{2P}\sum_{i=1}^{P}\left|\widehat{O_\iota} - O_i\right|^2 \tag{31}$$

The hyperparameter combination of the SGD controller and ReLU activation function is selected to train the DNN model. The regression curve obtained is depicted in Fig. 4, with the correlation coefficient $R$ reaching 0.998, which is quite satisfactory. The training process graph displayed in Fig. 5 indicates that the optimal model performance is attained at 260 epochs. Therefore, the number of epochs is chosen to be 260.

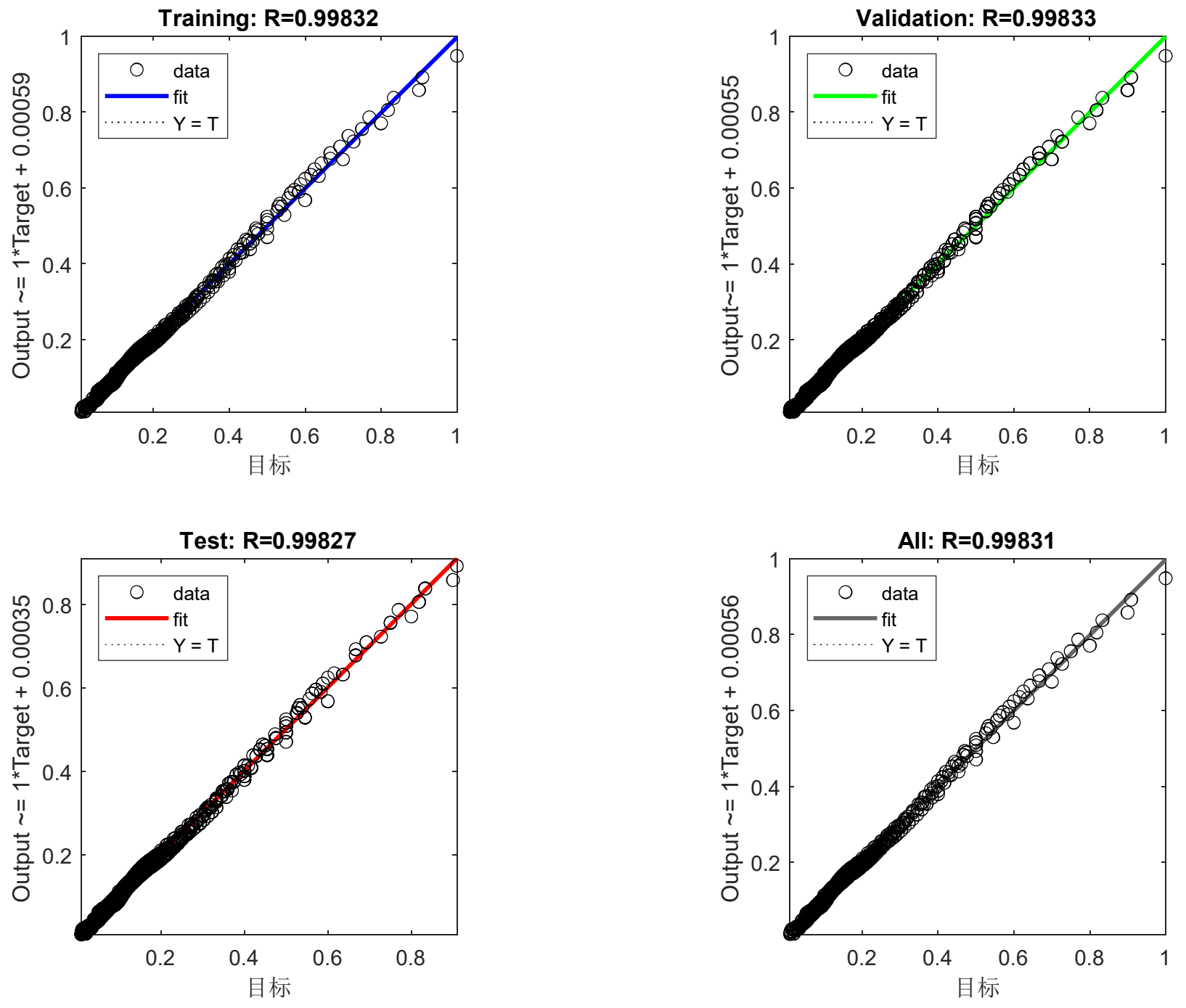

Fig. 4 Regression curve

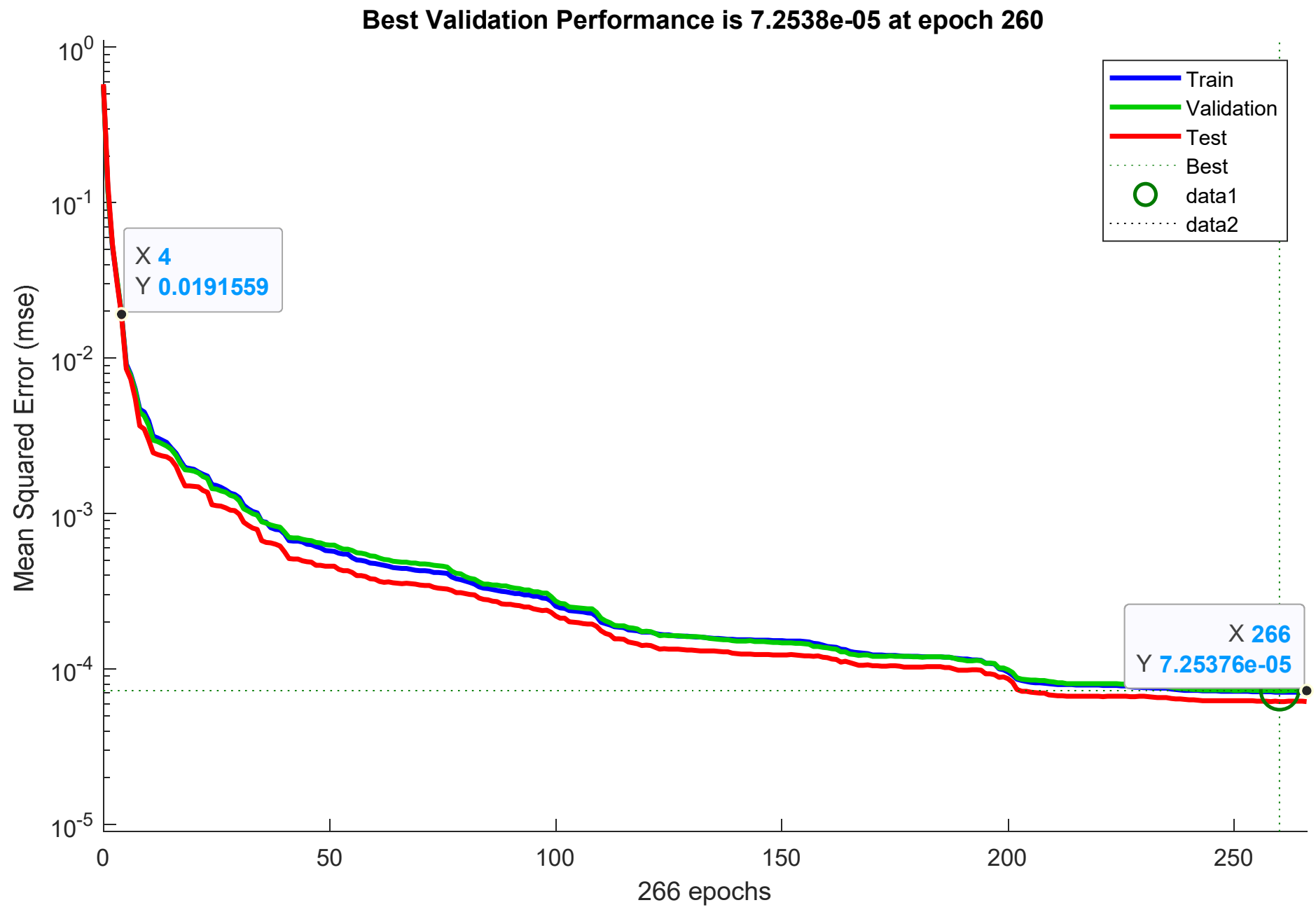

Figure 5 Training process curve

Arbitrary nonlinear second-order systems are represented as:

$$\begin{cases} \dot{x}_1 = x_2 \\ \dot{x}_2 = f(x) + g(x)u + d(t) \end{cases} \tag{32}$$

where $d(t)$ denotes the external disturbance to the system, satisfying $|d(t)| < D$, with $D$ being an upper limit value. $u$ is a control input, while $f(x)$ and $g(x)$ represent essential nonlinear functions defined by the actual mathematical model. The actual environment can be regarded as an uncertainty term. To mitigate the switching gain of the controller and address parameter perturbation uncertainty, the function $f(x)$ is approximated using a DNN. The adaptive sliding mode control system with DNN under closed-loop conditions is illustrated in Fig. 6.

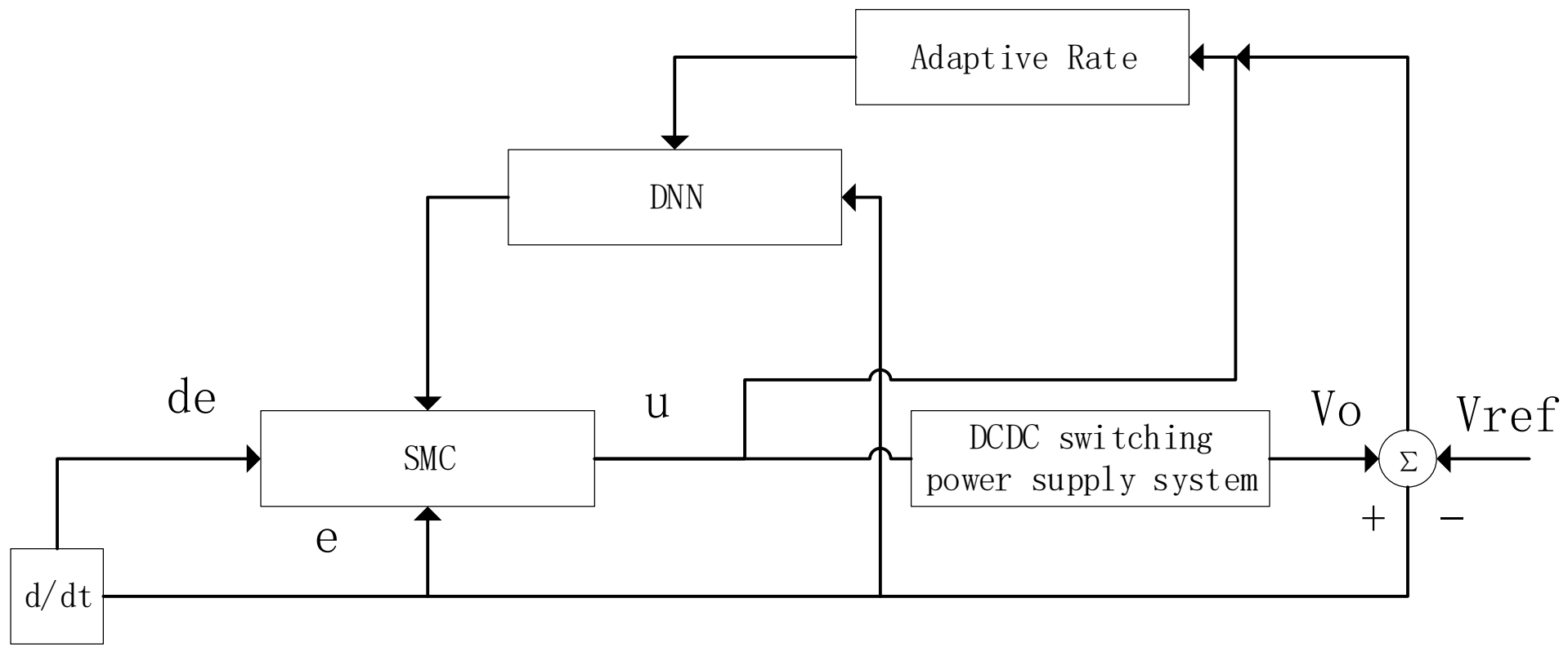

Fig. 6 Adaptive DNN-based sliding mode control system

In the DNN, the inputs are the tracking error $e$ and its rate of change $\dot{e}$, with the input function defined as $\boldsymbol{x} = [e\ \dot{e}]^T$. the fitted output of the network obtained by the equation $\hat{f}(\boldsymbol{x})$ is:

$$\hat{f}(\boldsymbol{x}) = \widehat{\boldsymbol{W}}^T \sigma(\boldsymbol{x}) \tag{33}$$

where $\widehat{\boldsymbol{W}}^T$ represents the estimated ideal network weights $\boldsymbol{W}^*$ of the estimation.

Let the error term of the fitted system be denoted as $\tilde{f} = f - \hat{f}$, so that:

$$\tilde{f} = -\widetilde{\boldsymbol{W}}^T \sigma(\boldsymbol{x}) + \varepsilon \tag{34}$$

where take $\widetilde{\boldsymbol{W}}^{*T} \sigma(\boldsymbol{x}) = \widehat{\boldsymbol{W}}^T \sigma(\boldsymbol{x}) - \boldsymbol{W}^{*T} \sigma(\boldsymbol{x})$.

From Eq. (20), introducing external disturbances:

$$\dot{s} = cx_2 - \frac{V_o}{LC} - \frac{\dot{V}_o}{RC} - \frac{V_i}{LC} D + d(t) \tag{35}$$

The expression for the corresponding term of the fit as:

$$f(x) = \frac{V_o}{LC} + \frac{\dot{V}_o}{RC} \tag{36}$$

Take the control rate as:

$$u = \frac{LC}{V_i} [-cx_2 - \eta\ sgn(s) - \hat{f}(\boldsymbol{x})] \tag{37}$$

Design the Lyapunov function as:

$$V = \frac{1}{2}s^2 + \frac{1}{2\gamma}\widetilde{\boldsymbol{W}}^T\widetilde{\boldsymbol{W}} \tag{38}$$

Take the derivative:

$$\begin{aligned}
\dot{V} &= s\dot{s} + \frac{1}{\gamma}\widetilde{\boldsymbol{W}}^T\dot{\widehat{\boldsymbol{W}}} \\
&= s\left[cx_2 + (\frac{uV_i}{LC} + f + d(t))\right] + \frac{1}{\gamma}\widetilde{\boldsymbol{W}}^T\dot{\widehat{\boldsymbol{W}}} \\
&= s[-\hat{f} + f + d(t) - \eta\, sgn(s)] + \frac{1}{\gamma}\widetilde{\boldsymbol{W}}^T\dot{\widehat{\boldsymbol{W}}} \\
&= s[\tilde{f} + d(t) - \eta\, sgn(s)] + \frac{1}{\gamma}\widetilde{\boldsymbol{W}}^T\dot{\widehat{\boldsymbol{W}}} \\
&= s[-\widetilde{\boldsymbol{W}}^T\sigma(\boldsymbol{x}) + \varepsilon - \eta\, sgn(s) + d(t)] + \frac{1}{\gamma}\widetilde{\boldsymbol{W}}^T\dot{\widehat{\boldsymbol{W}}} \\
&= \widetilde{\boldsymbol{W}}^T\left(-\sigma(\boldsymbol{x})s + \frac{1}{\gamma}\dot{\widehat{\boldsymbol{W}}}\right) + \varepsilon s - \eta|s| + T|s|
\end{aligned} \tag{39}$$

where $T$ is the upper bound of $d(t)$. The law of adaptation is taken as:

$$\dot{\widehat{\boldsymbol{W}}} = \gamma s\sigma(\boldsymbol{x}) \tag{40}$$

Then:

$$\begin{aligned}
\dot{V} &\leq \varepsilon s - \eta|s| + T|s| \\
&\leq (-\eta + \varepsilon_N + T)|s| \\
&\leq 0
\end{aligned} \tag{41}$$

Based on the above equations, it is evident that the designed DC-DC buck converter system with DNN-based sliding mode control satisfies Lyapunov's theorem, thus demonstrating the stability of the resulting system.

# 4. Results and discussions

*4.1. Simulation results*

To verify the feasibility of the designed converter, we build an analog simulation platform using Simulink in MATLAB for experimental analysis. The main parameters of the DC-DC buck switching power supply are as follows: a $12V$ input voltage $V_{in}$, a $5V$ output voltage $V_{out}$, an inductance $L$ of $160uH$, a capacitance $C$ of $200uF$, a switching frequency $f_s$ of $25kHz$, and load resistances $R_L$ of $2\Omega$ and $10\Omega$. In the circuit, the load abruptly changes from $2\Omega$ to $10\Omega$ within 0.03 seconds. The Simulink circuit simulation is depicted in Fig. 7.

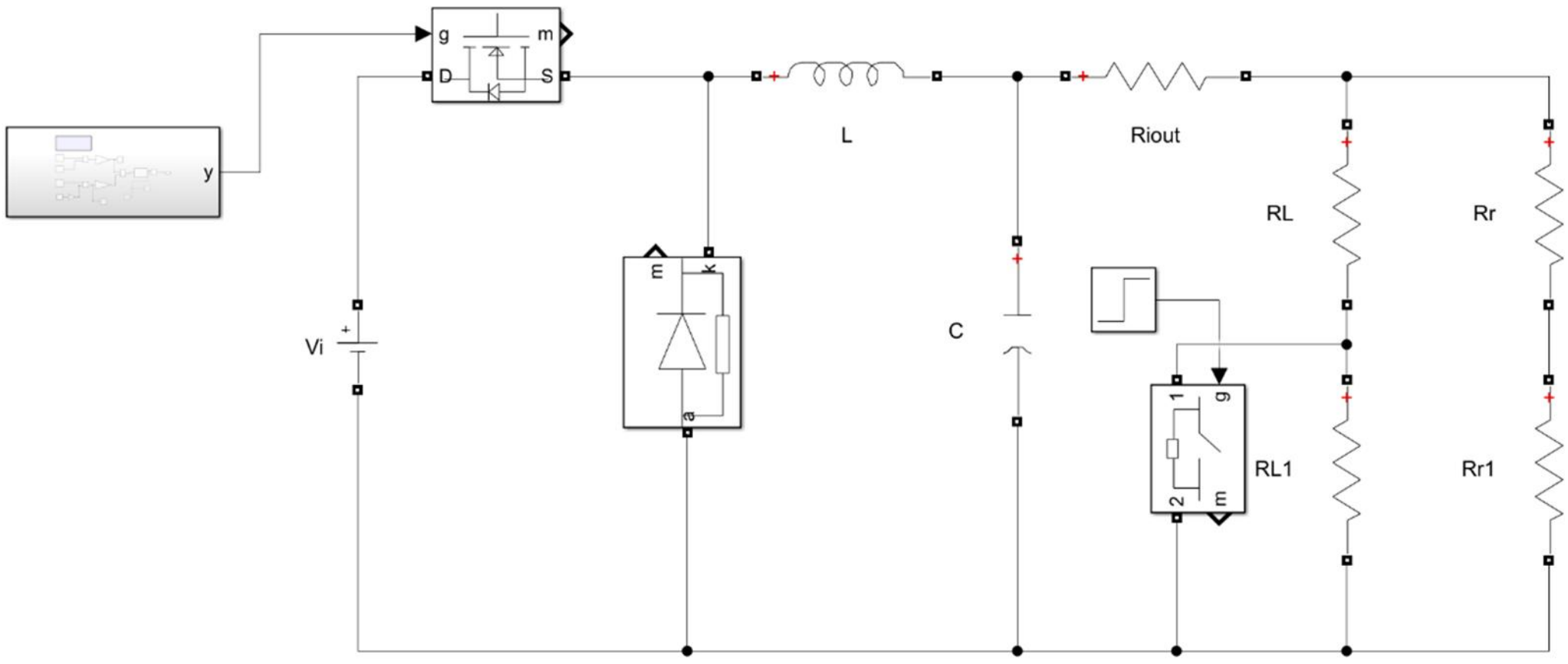

Fig. 7 Simulink simulation of a DC-DC buck circuit.

For classic sliding mode control, the inductor current simulation waveform is shown on the left of Fig. 8, and the output voltage simulation waveform is shown on the left of Fig. 9, with a target voltage of $5V$.

The figure indicates that the voltage stabilizes at approximately $4.991V$ around 0.03 seconds. At this point, the load abruptly changes from $2\Omega$ to $10\Omega$. Consequently, (1) the inductor current surges from $0.5A$ to $2.5A$, and the system requires approximately 9ms to stabilize the current. (2) The output voltage exhibits fluctuations, and although the system eventually regulates the output voltage within 40ms and stabilizes it, chattering persists.

For DNN-based sliding mode control, the inductor current simulation waveform is shown on the right of Fig. 8, and the output voltage simulation waveform is shown on the right of Fig. 9. At 0.03 seconds, the inductor current undergoes a jump from 0.5A to 2.5A. Comparatively, the DNN-based sliding mode control results in a significantly reduced fluctuation in the inductor current, stabilizing in approximately 600μs, whereas classical sliding mode control may exhibit a longer stabilization time. Figure 7 illustrates that the voltage reaches the target voltage within approximately 8ms, which is roughly twice the convergence speed achieved with classical sliding mode control. At 0.03 seconds, the load transitions from 2Ω to 10Ω. During this abrupt change, the output voltage exhibits fluctuations; however, the system successfully completes voltage regulation within 20ms. Remarkably, the system stabilizes within 0.03 seconds, showcasing its enhanced resilience to interference compared to classical sliding mode control. Notably, the DNN-based sliding mode control effectively eliminates chattering.

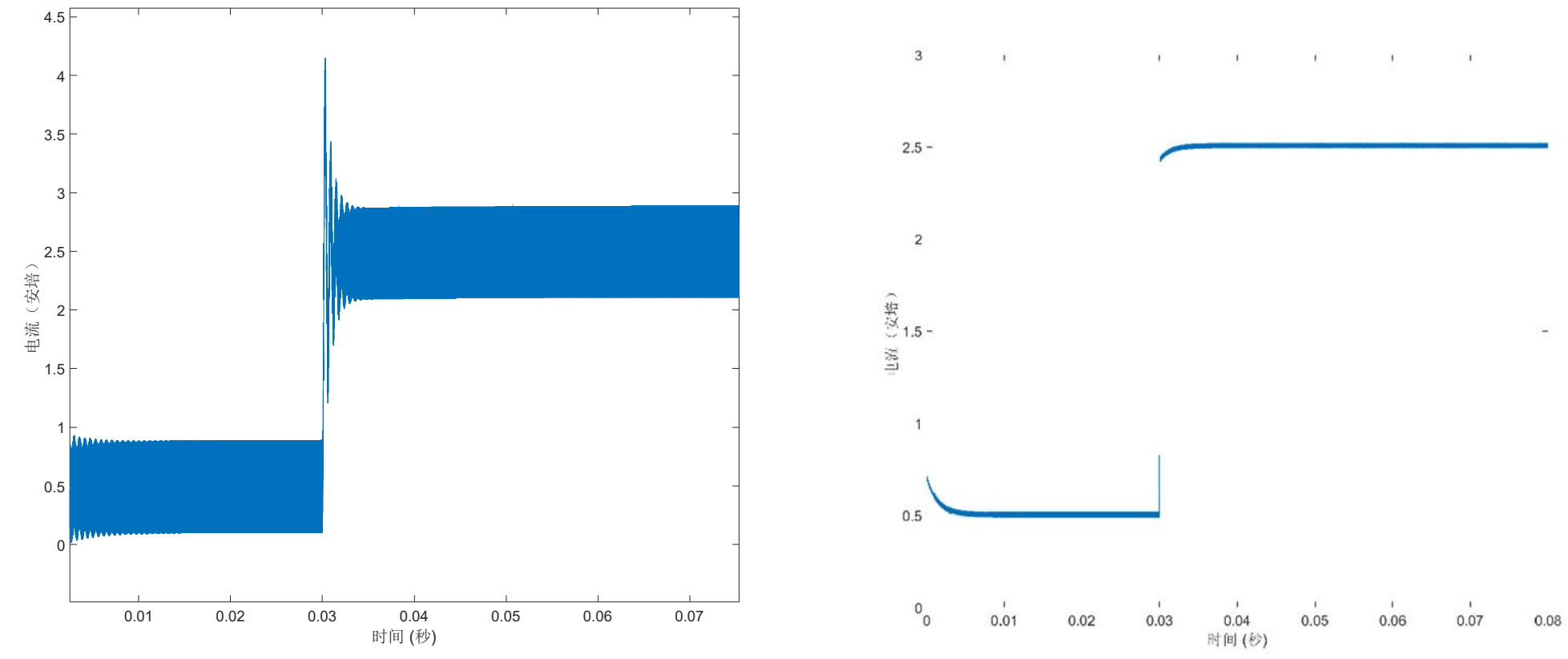

Fig. 8 Simulated waveforms of inductor currents under classic sliding mode control and DNN-based sliding mode control.

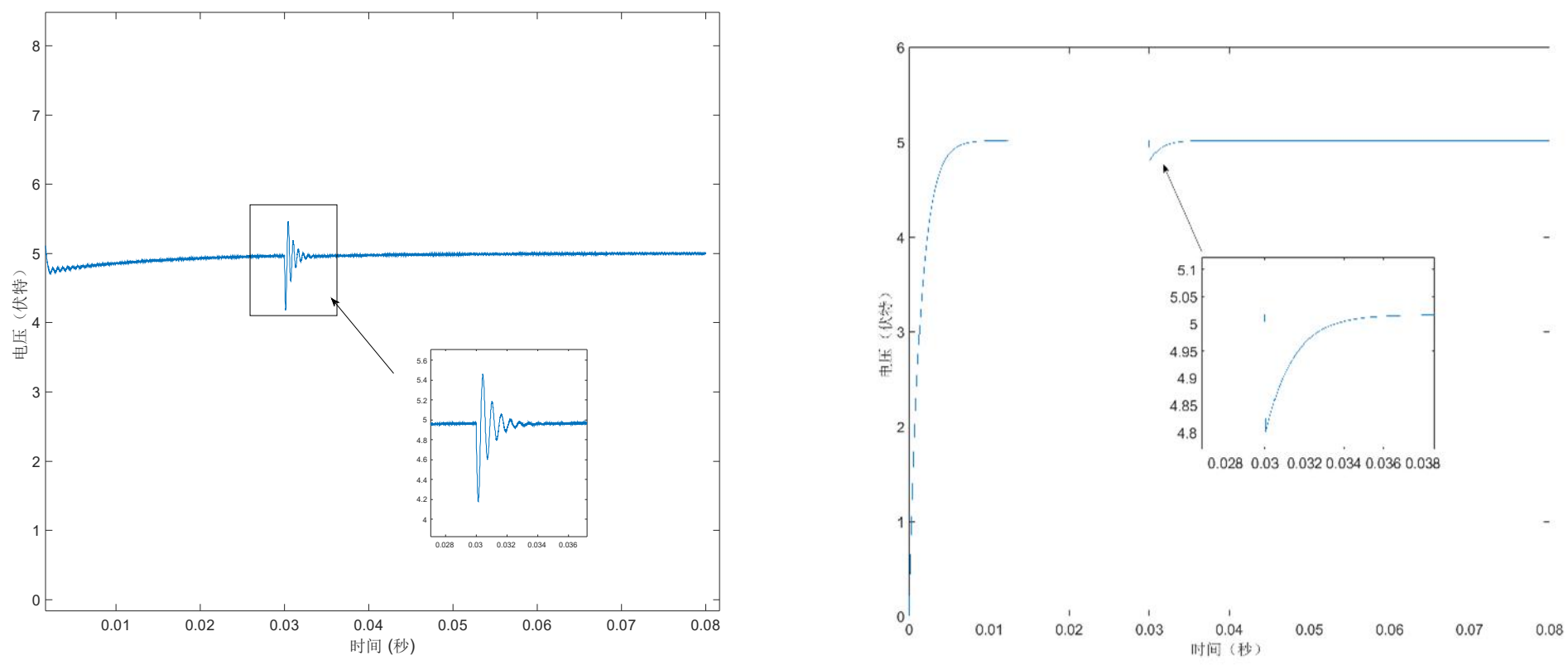

Fig. 9 Simulated waveforms of output voltage under classic sliding mode control and DNN-based sliding mode control.

### *4.2. Experimental results*

Fig. 10 shows the hardware experimental setup. The experimental platform of the DC-DC converter includes the buck-type DC-DC main circuit, an auxiliary power supply, an MCU controller, and a software component implemented in PSIM. The control model of the DC-DC converter is built in the PSIM software. Then, using JTAG tools, the circuit model is converted to C code and burned into the MCU control circuit module. After the circuit starts, the output voltage and inductor current are fed back to the MCU chip through the voltage-current sampling circuit. The MCU, utilizing its powerful computational capabilities, calculates the output value of the DNN-based sliding mode control strategy. Finally, the PWM waveform controlling the buck circuit is generated.

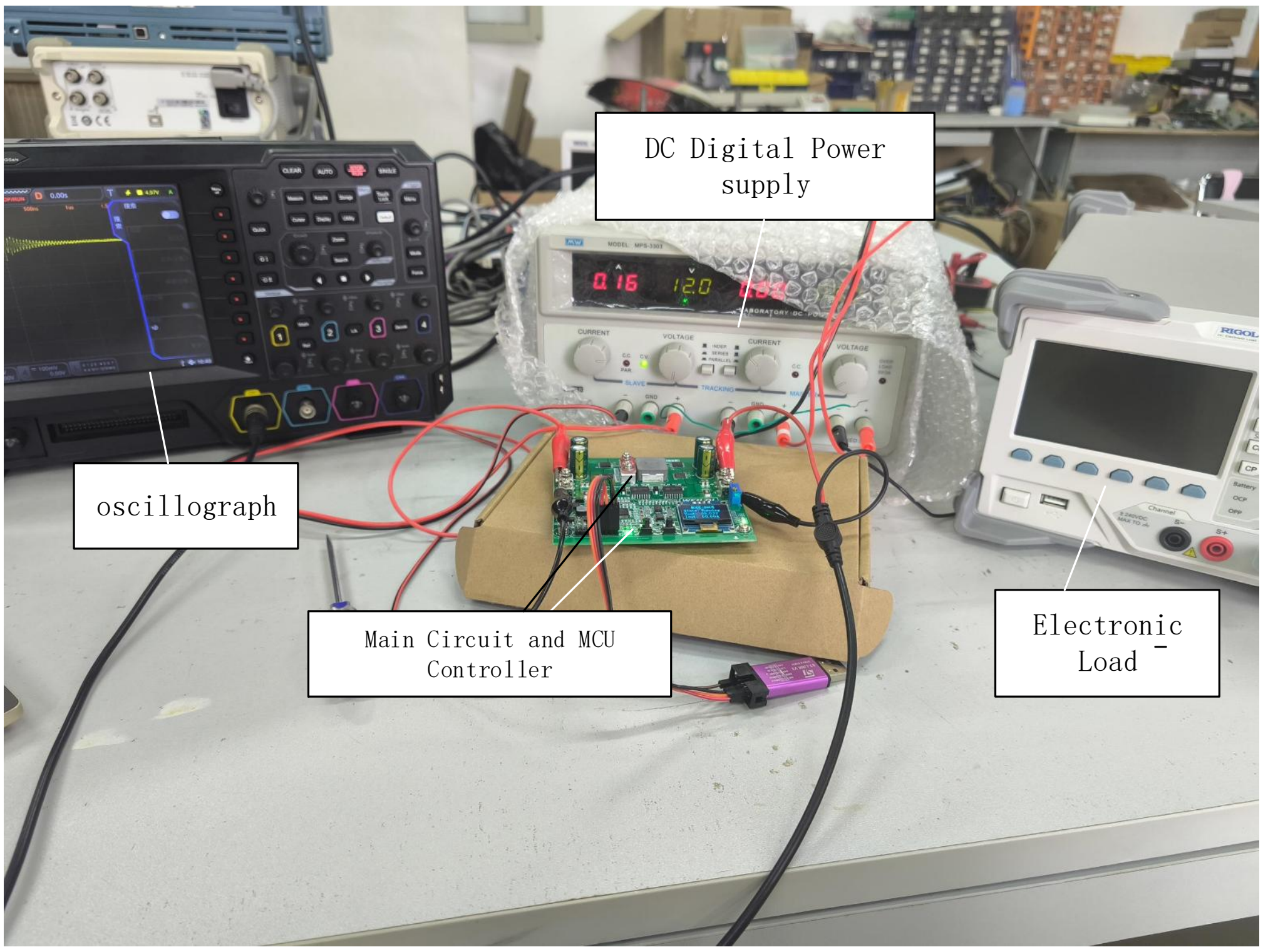

Fig. 10 Hardware experiment platform

The MCU controller used in this paper is the STM32F334 chip, which is characterized by strong scalability and flexible operation modes, making it very suitable for controlling buck circuits.

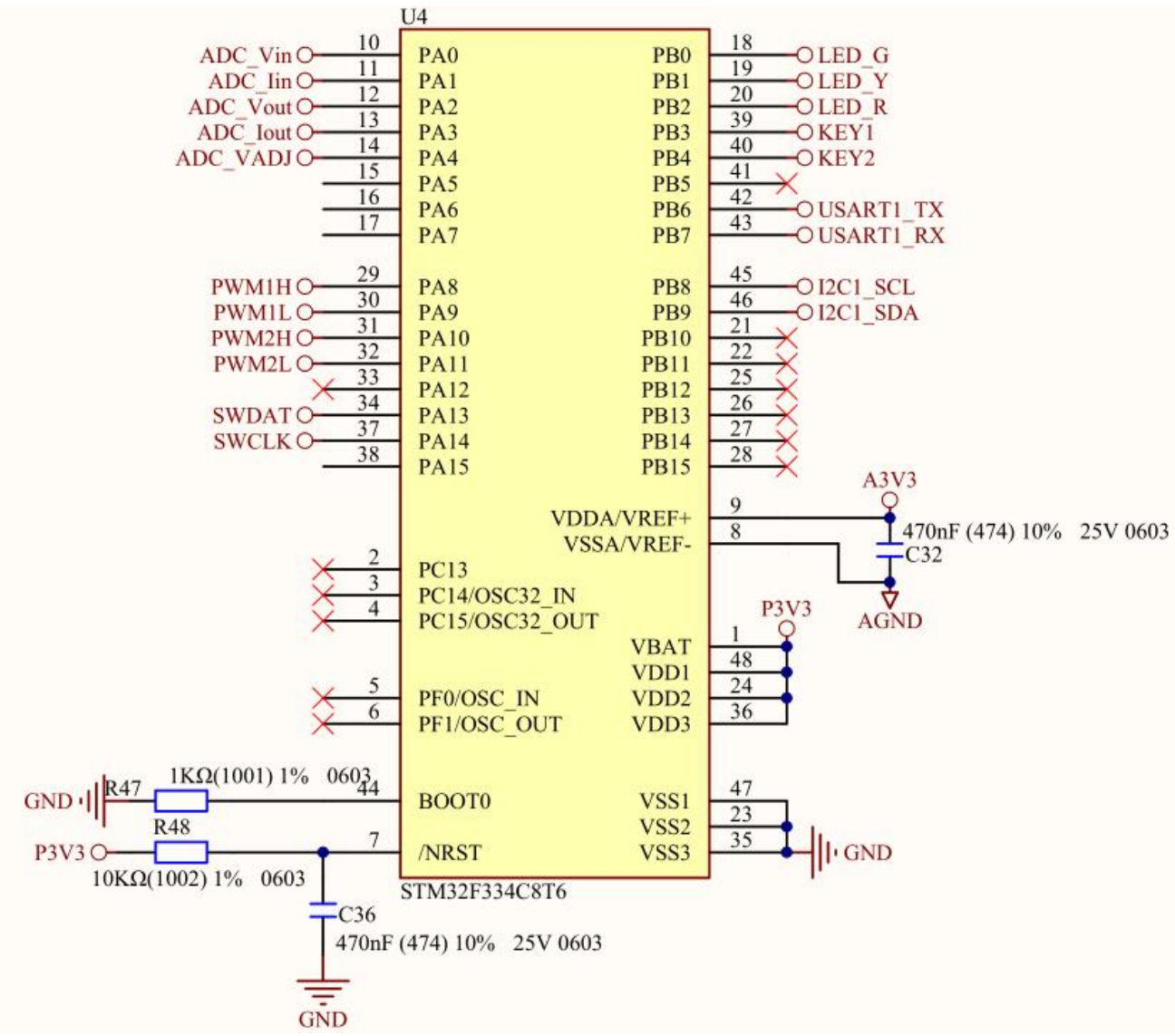

Fig. 11 MCU control circuit

The DNN controller algorithm needs to be implemented on the MCU controller platform. First, various datasets were carefully constructed using DC-DC converters under different operating conditions. These datasets were then saved in Comma Separated Values (CSV) format for convenient storage in Excel files. The DNN model was trained using the neural network model in MATLAB 2022a software. Upon completion of the training phase, the trained DNN model was deployed to the MCU controller via PSIM software using the MATLAB support package. This deployment method enables the controller to run autonomously in an independent mode, ensuring efficient real-time control performance. Table 1 shows the main parameters selected for the DNN.

Table 2 Parameters of the DNN

| **Parameters** | **DNN** |
|---|---|
| Weight update rule | SGD |
| Performance metric | RMSE |
| Epochs | 50 |
| Activation function | ReLU |
| No. of input nodes | 2 |
| No. of hidden layer1 nodes | 3 |
| No. of hidden layer2 nodes | 3 |
| No. of output layer nodes | 1 |

The main parameters of the DC-DC buck switching power supply are respectively: $12V$ The input voltage of $V_{in}$ the $5V$ The reference output voltage of $V_{out}$ the reference output voltage $160uH$ the inductance $L$ the $200uF$ the capacitance of the$C$ the $25khz$ The switching frequency of $f_s$ of the inductor$10\Omega$ The load resistance of $R_L$ of the load resistance.

In this paper, three sets of controlled experiments are conducted to compare the experimental data of the DC-DC buck switching power supply using classical sliding mode control and the DC-DC buck switching power supply using DNN-based sliding mode control:

(1) Set the rated input voltage to $12V$ and the load resistance to $10\Omega$ , and observe the output voltage response in both sets of experiments.

(2) After the output voltage has stabilized, change the load resistance from the original $10\Omega$ to $2\Omega$, and observe the system's response to load disturbance.

(3) After the output voltage has stabilized, change the input voltage from the original $12V$ to $13V$, and observe the system's response to voltage disturbance.

Experiment 1: Fig. 12 shows the observed output voltage response of the classical sliding mode controller and the DNN-based sliding mode controller. It can be concluded from the observation that the response speed of the classical sliding mode controller is $29.8\ ms$, while the response speed of the DNN-based sliding mode controller can reach $9.7\ ms$. After the system stabilizes, the DNN-based sliding mode control also has fewer burrs and better ripple suppression than the classical sliding mode control.

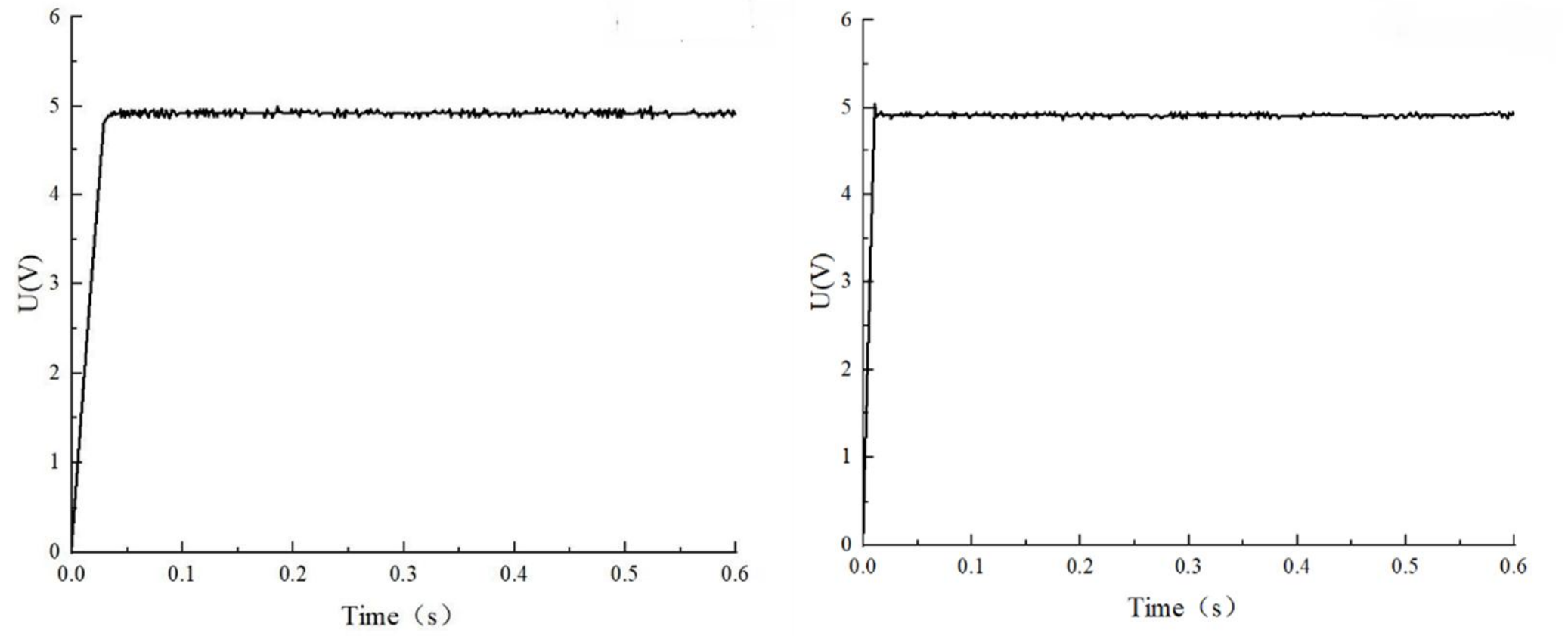

(a) Classical sliding mode controller (b) DNN-based sliding mode controller

Fig. 12 Output voltage response of the controller

Experiment 2: Simulation of load disturbances. After the system stabilizes at $12V$, the load resistance is manually changed by adjusting the value of the electronic resistor to test the system's ability to resist load interference.

After the circuit has stabilized at $12V$, the load resistance is modified from $10\Omega$ to $2\Omega$. The output response graph is shown in Fig. 13.

Fig. 13 shows the observed output voltage response of the classical sliding mode controller and the DNN-based sliding mode controller after the change in load resistance. It can be observed that the overshoot of the classical sliding mode controller is $0.25V$, while the overshoot of the DNN-based sliding mode controller is $0.24V$, both are similar. In terms of recovery time, the latter is superior to the former, with the classical sliding mode controller having a recovery time of $9.6ms$, while the DNN-based sliding mode controller has a recovery time of $4.8ms$ Compared with the classical sliding mode controller, the DNN-based sliding mode controller has a reduced recovery time of $50\%$.

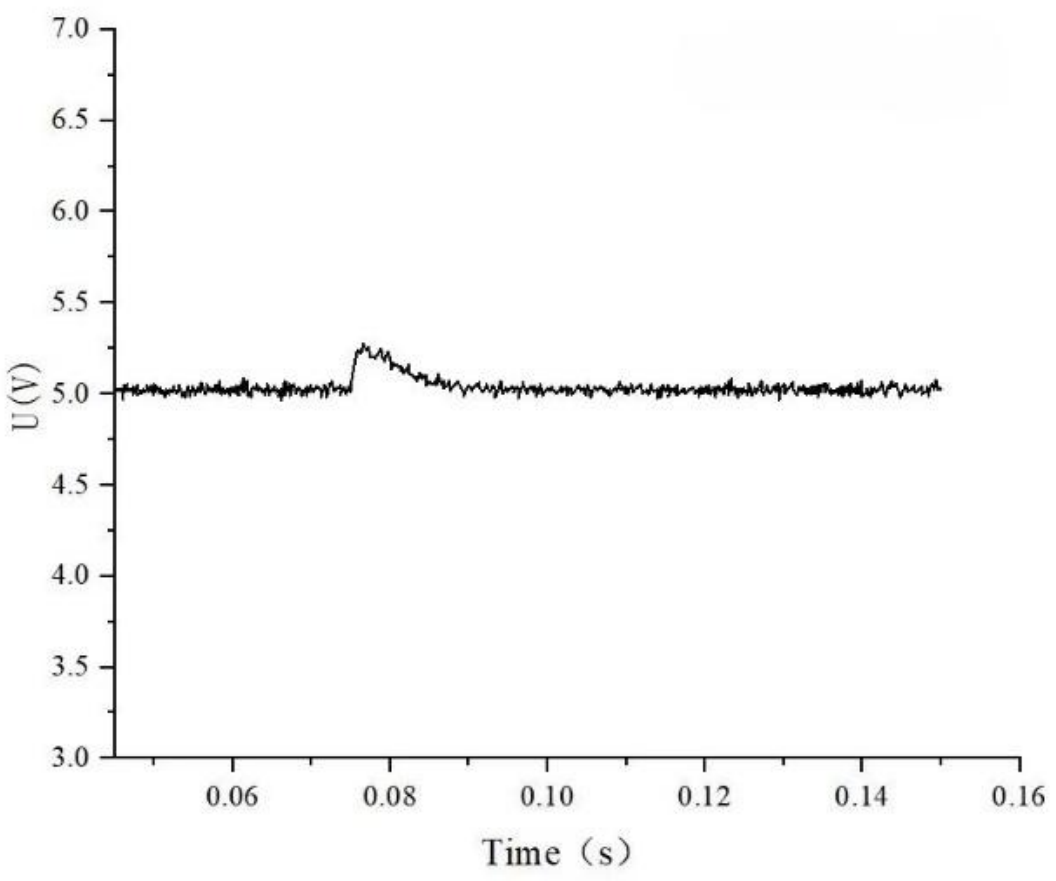

(a) Classical sliding mode controller

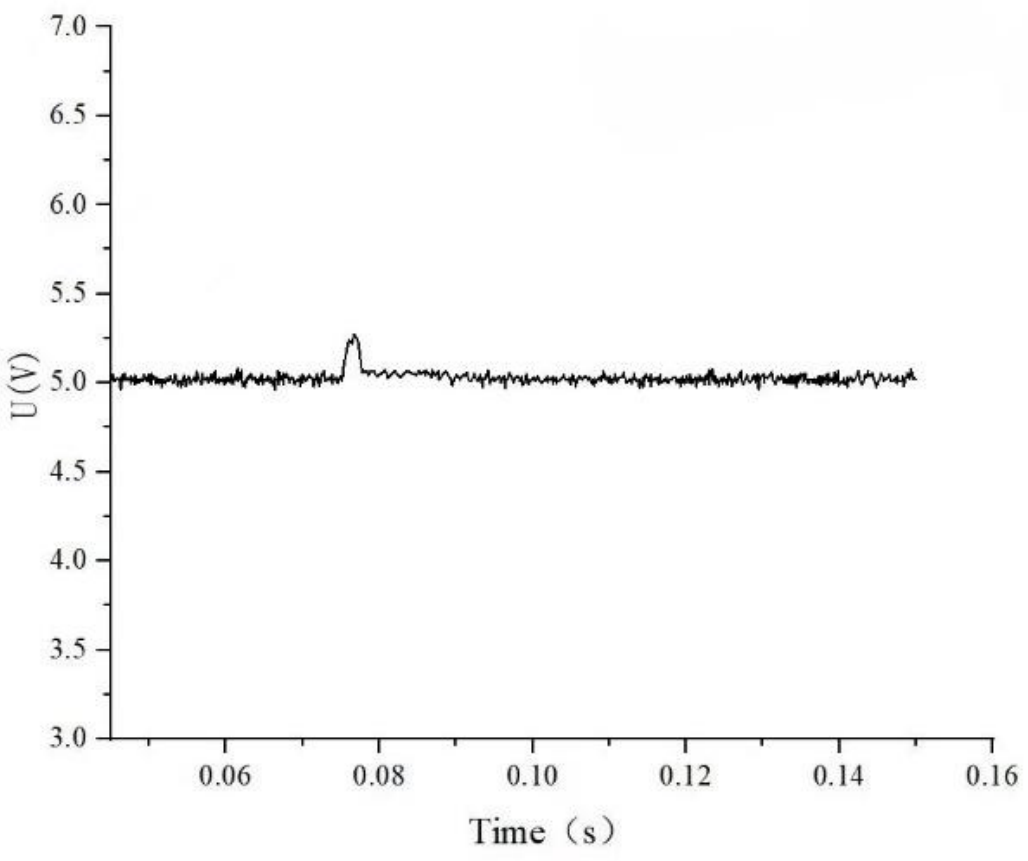

(b) DNN-based sliding mode controller

Fig. 13 Response graph of load resistance change of the controller

Experiment 3: Simulation of voltage interference. This experiment tests the system's resilience to voltage interference by manually adjusting the input voltage of the DC power supply.

After the circuit has stabilized at $12V$, the input voltage is then increased from $12V$ to $13V$,.The output response graph is shown in Fig. 14.

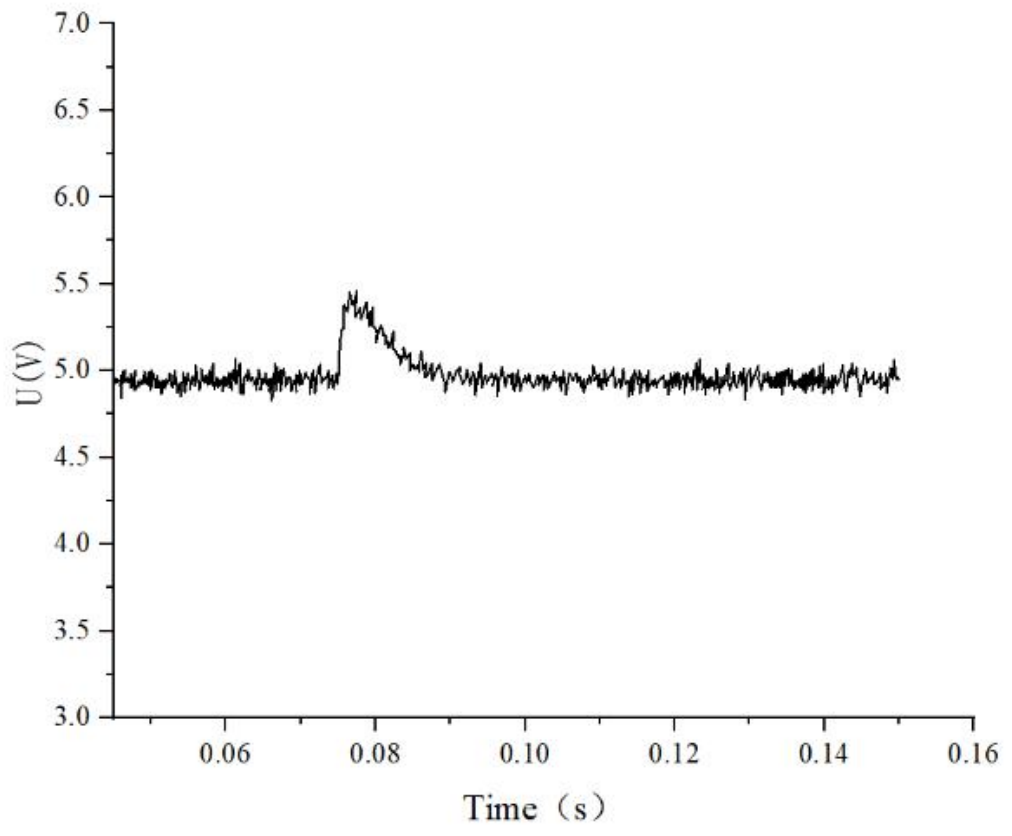

(a) Classical sliding mode controller

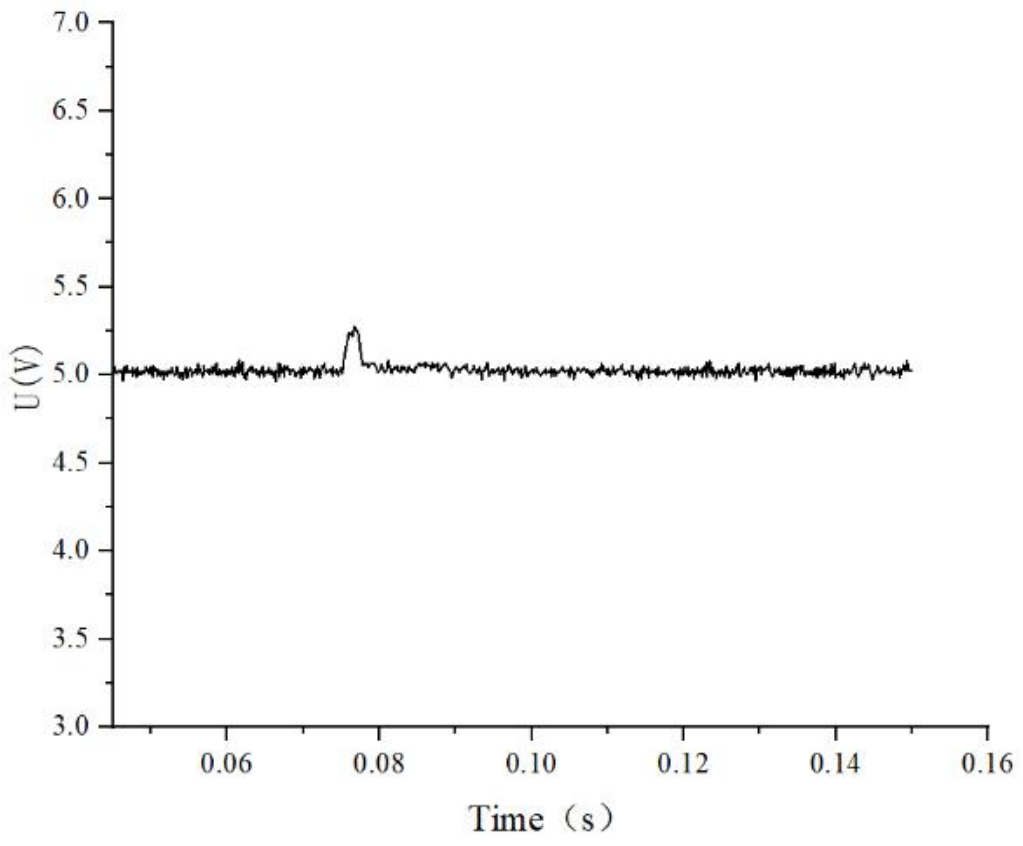

(b) DNN-based sliding mode controller

Fig. 14 Input voltage change response graph of the controller

Fig. 14 shows the observed output voltage response of both the classical sliding mode controller and the DNN-based sliding mode controller after the input voltage is altered. Observations indicate a notable difference in overshoot between the two controllers: the classical sliding mode controller exhibits a peak overshoot of up to $0.48V$, while the DNN-based sliding mode controller's overshoot is limited to $0.23V$. This suggests that the latter possesses superior resilience to input voltage disturbances. Additionally, considering recovery time, the DNN-based sliding mode controller outperforms the classical sliding mode controller. Specifically, the recovery time for the classical sliding mode controller is approximately 12.1ms, while for the DNN-based sliding mode controller, it is only about 5.1ms, representing a 58% reduction. This emphasizes the DNN-based controller's faster response to disturbances. Overall, the experiment highlights that input voltage disturbances have a more pronounced impact on buck circuits compared to load disturbances, corroborating the general trend.

The above experimental results are organized in Table 3. Due to non-ideal factors in the hardware circuits, such as circuit delay, electromagnetic interference, or the response rate of the optoelectronic devices, there may still be a certain gap between the experimental results and the simulation results. However, upon comparing the experiments, it is evident that the DNN-based sliding mode controller outperforms the classical sliding mode controller in several aspects. Specifically, the DNN-based controller exhibits significantly reduced chattering, faster response times, and superior anti-interference capabilities compared to the classical sliding mode controller.

Table 3 Experimental result data

| | Output response | | Load disturbance | | Input voltage disturbance | |
|---|---|---|---|---|---|---|
| Control strategy | Voltage ripple | Response time （$ms$） | Overshoot voltage （$V$） | Recovery time （$ms$） | Overshoot voltage （$V$） | Recovery time （$ms$) |
| SMC | High | 29.8 | 0.25 | 9.6 | 0.48 | 12.1 |
| DNN-SMC | Low | 9.7 | 0.24 | 4.8 | 0.23 | 5.1 |

# 5 Conclusion

Aiming to improve convergence speed, reduce chattering, and enhance robustness，a DNN-based sliding mode control design for a DC-DC buck switching power supply has been proposed in this paper. Compared with the classical sliding mode control, the strategy in this paper has better performance in terms of convergence speed, chatter suppression and robustness. This study introduces a unique control strategy for DC-DC converters, offering a fresh perspective for future designs.